\documentclass[aps,prl,twocolumn,groupedaddress]{revtex4}
\usepackage{graphicx}
\usepackage{amsmath}

\begin{document}

\title{Quantum annealing with Jarzynski equality}
\author{Masayuki Ohzeki}
\affiliation{Department of Systems Science, Graduate School of Informatics, Kyoto
University, 36-1 Yoshida-Honmachi, Sakyo-ku, Kyoto, 606-8501, Japan}
\date{\today}

\begin{abstract}
We show a practical application of the Jarzynski equality in quantum
computation. Its implementation may open a way to solve combinatorial
optimization problems, minimization of a real single-valued function, cost
function, with many arguments. We consider to incorpolate the Jarzynski
equality into quantum annealing, which is one of the generic algorithms to
solve the combinatorial optimization problem. The ordinary quantum annealing
suffers from non-adiabatic transitions whose rate is characterized by the
minimum energy gap $\Delta$ of the quantum system under consideration. The
quantum sweep speed is therefore restricted to be extremely slow for the
achievement to obtain a solution without relevant errors. 
However, in our strategy shown in the present study, we find that such a difficulty would not matter.
\end{abstract}

\maketitle

Quantum computer is believed to be able to solve intractable problems on
classical computer by use of superposition, tunneling effect and
entanglement attributed to quantum nature. This fascinating device is
attracted to researchers and studied from both aspects of the fundamental
interests of quantum nature and its application. The intractable problems we
wish to solve are often closely related to efficiency for cost and time in
industry and distribution systems to gain convenience in our daily life. 
The efforts in both of the theoretical and experimental approaches have culminated to realize quantum computer, which can mitigate the difficulties to solve such hard problems. 
Hundreds of hard problems are found in optimization problem, which is a kind of the problems to minimize or maximize a real single-valued function of multivariables called the cost function.
The cases
in which variables take discrete values are known as combinatorial optimization, whose well-known instances are satisfiability problems, exact
cover, maximum cut, Hamilton graph, and traveling salesman problem \cite{OP,OP2}. 
Most of the interesting optimization problems belong to the hard class in which the best known algorithms cost exponentially long time as a function of the system size (the number of degrees of freedom representing the cost function). 
Therefore one desires quantum computation, which enables us to solve such hard optimization problems by an algorithm employing quantum nature. 
One of the generic algorithms proposed as part of such efforts is quantum annealing (QA) \cite{QA1,QA2,QA3}. 
In QA, we introduce artificial degrees of freedom of quantum nature, noncommutative operators, which induce quantum fluctuations to drive the system as 
\begin{equation}
H(t)=f(t)H_{0}+\left\{ 1-f(t)\right\} H_{1},  \label{QA}
\end{equation}%
where $H_{0}$ is the classical Hamiltonian consisting of diagonal elements,
which express the cost function. Here $f(t)$ is assumed to be a
monotonically increasing function satisfying $f(0)=0$ and $f(\tau )=1$. The
quantum annealing starts from a single pure state, the ground state of $H_{1}
$, which is chosen to be trivially given as $|\Psi (0)\rangle=\sum_{\{\sigma \}}|\sigma \rangle /\sqrt{N}$, where $N$ characterizes the system size of the optimization problem. 
The adiabatic theorem guarantees that we can reach a nontrivial ground state of $H_{0}$ after quantum dynamics with sufficiently slow speed as $1/\tau _{c}\sim \Delta_{\rm min.}^{2}$, where $\tau $ means the annealing time, and $\Delta $ is the energy gap of
the instantaneous quantum system as in Eq. (\ref{QA}) \cite{QA3,AD}. 
However QA does not work well in a reasonable time, since we require extremely slow control, for the cases in which the quantum system as in Eq. (\ref{QA}) has a minimum energy gap vanishing as $\Delta_{\rm min.} \sim \exp(-\alpha N)$ for increasing the system size $N$ \cite{FT,FT3}. 
The quantum annealing is a very generic technique but has such a bottleneck.

To overcome the above difficulty, we bring another theoretical piece from
non-equilibrium statistical physics, the Jarzynski equality (JE), in the
present study \cite{JE1,JE2}. The Jarzynski equality is written by an
well-known expression as, 
\begin{equation}
\left\langle \mathrm{e}^{-\beta W}\right\rangle =\frac{Z_{\tau }(\beta )}{%
Z_{0}(\beta )},
\end{equation}%
where the angular brackets denote the average over all realizations in a
predetermined process starting from an initial equilibrium state and $W$ is
the work done during the process. The partition functions for the initial
and final Hamiltonians are written as $Z_{0}(\beta )$ and $Z_{\tau }(\beta )$
with inverse temperature $\beta $, respectively. We here recall the formulation of JE for classical systems on a heat bath \cite{JE2}. 
Let us consider a thermal nonequilibrium process in a finite-time schedule $0\leq t\leq \tau $. 
Thermal fluctuations can be simulated by the master equation. 
We employ discrete time expressions and write $t_{k+1}-t_{k}=\delta t$, $t_{0}=0$ and $t_{n}=\tau $. 
The probability that the system is in a state $\sigma _{k}$ at time $t_{k}$ is denoted as $P(\sigma _{k};t_{k})$. 
The transition probability per unit time $\delta t$ is defined as $M(\sigma
_{k+1}|\sigma _{k};t_{k})$. In the original formulation of JE, the work is
defined as the energy difference merely attributed to the change of the Hamiltonian, but we can construct JE also in the case of changing the inverse temperature by defining the work as $-\beta W(\sigma_{k};t_{k})=-(\beta (t_{k+1})-\beta (t_{k}))E(\sigma _{k})$, where $E(\sigma)$ is the value of the cost function (classical Hamiltonian $H_{0}$) for the specific state $\sigma $. 
The left-hand side of JE can be expressed as 
\begin{eqnarray}
\left\langle \mathrm{e}^{-\beta W}\right\rangle  &=&\sum_{\{\sigma
_{k}\}}\prod_{k=0}^{n-1}\left\{ \mathrm{e}^{-\beta W(\sigma _{k+1};t_{k})}%
\mathrm{e}^{\delta tM(\sigma _{k+1}|\sigma _{k};t_{k})}\right\}   \notag \\
&&\quad \times \tilde{P}(\sigma _{0};t_{0}),  \label{CJE}
\end{eqnarray}%
where $\tilde{P}(\sigma _{0};t_{0})$ denotes the initial equilibrium
distribution. The initial condition is set to the equilibrium distribution.
If the transition term $\exp (\delta tM(\sigma _{k+1}|\sigma _{k};t_{k}))$ is removed in Eq. (\ref{CJE}), JE is trivially satisfied because the summation
of $-\beta W(\sigma _{k+1};t_{k})$ over $k$ yields $-(\beta (t_{n})-\beta
(t_{0}))E(\sigma _{0})$. A non-trivial aspect of JE is in the insertion of
the transition term, which does not alter the conclusion. From Eq. (\ref{CJE}), it is straightforward to prove JE. 
This is the case for classical systems on a heat bath, not for quantum systems. 
One may think the above classical equality is not available for the application to QA. 
Nevertheless we can apply the classical JE to QA by aid of the classical-quantum mapping \cite{QC}.

The classical-quantum mapping leads us to a special quantum system, in which the (instantaneous) equilibrium state of the above stochastic dynamics can be expressed as a ground state. 
A general form of such a special quantum Hamiltonian is 
\begin{equation}
H_q(\sigma^{\prime }|\sigma;t) = \delta_{\sigma^{\prime },\sigma} - \mathrm{e%
}^{\beta(t)H_0(\sigma^{\prime })/2}M(\sigma^{\prime }| \sigma;t) \mathrm{e}%
^{-\beta(t)H_0(\sigma)/2}.  \label{Hq}
\end{equation}
This Hamiltonian has the ground state as $|\Psi_{\mathrm{eq}%
}(t)\rangle=\sum_{\sigma }\mathrm{e}^{-\frac{\beta (t)}{2}%
H_{0}(\sigma)}|\sigma\rangle/\sqrt{Z(t)}$. 
It is clear that the quantum
expectation value of a physical quantity $A(\sigma)$ by $|\Psi_{\mathrm{eq}%
}(t)\rangle$ is equal to the thermal expectation value for the same
quantity. The ground state energy is $0$, which can be explicitly shown by
the detailed-balance condition.
On the other hand, the excited states have positive-definite eigenvalues, which can be confirmed by the Perron-Frobenius theorem.

In the above special quantum system, we can treat a quasi-equilibrium
stochastic process as an adiabatic quantum-mechanical dynamics in QA. Let us
consider QA for the above special quantum system by setting the parameter
corresponding to the temperature $T\to \infty$ ($\beta \to 0$). This
condition gives the trivial ground state with uniform linear combination,
similarly to the ordinary QA. If we tune $T\to 0$ very slowly, one can
obtain the ground state for $H_q$, which expresses the very low-temperature
equilibrium state for $H_0$, the cost function of the optimization problem
that we wish solve. Notice that we use a single quantum state during the
above procedure, not an ensemble assumed in the ordinary formulation in JE.

Let us construct a protocol with the same spirit as JE by using the special
quantum system. Initially we prepare the trivial ground state with the uniform
linear combination as in the ordinary QA. From the point of view of the
classical-quantum mapping, this initial state expresses the high-temperature
equilibrium state $|\Psi_{\mathrm{eq}} (t_0)\rangle \propto \mathrm{e}%
^{-\beta(t_{0})H_{0}(\sigma)/2}|\sigma \rangle $ with $\beta(t_0)\ll 1$. We
introduce the exponentiated work operator $W(\sigma_{k};t_k)=\exp
(-(\beta(t_{k+1}) -\beta (t_{k}))H_{0}(\sigma_k)/2)$. It looks like a
non-unitary operator, but we can construct this operation by considering an
extended quantum system as discussed later. If we apply $W(\sigma_{k};t_k)$
to the quantum wave function $|\Psi_{\mathrm{eq}} (t_k)\rangle$, the state
is changed into a state corresponding to the equilibrium distribution with
the inverse temperature $\beta (t_{k+1})$. When the time-evolution operator $%
U(\sigma^{\prime}|\sigma;t_{k+1})=\exp (-\mathrm{i}\delta t
H_{q}(\sigma^{\prime}|\sigma;t_{k+1})/\hbar)$ is applied, this state does
not change, since it is the ground state of $H_{q}(\sigma^{\prime}|%
\sigma;t_{k+1})$. The obtained state after the repetition of the above
procedure is 
\begin{eqnarray}
|\Psi (t_{n})\rangle & \propto & \prod_{k=0}^{n-1}\left\{
W(\sigma_{k+1};t_{k})U_{k+1}(\sigma_{k+1}|\sigma_{k};t_{k}) \right\}|\Psi_{%
\mathrm{eq}} (t_{0})\rangle .  \notag \\
\end{eqnarray}
This is essentially of the same form as Eq. (\ref{CJE}). Instead of the
exponentiated matrix of $\delta t M(\sigma_{k+1}|\sigma_{k};t_{k})$, we use the
time-evolution operator $U(\sigma_{k+1}|\sigma_{k};t_{k})$ here. After the
system reaches the state $|\Psi (t_n)\rangle$, we measure the obtained state
by the projection onto a specified state $\sigma ^{\prime }$. The
probability is then given by $|\langle \sigma^{ \prime }|\Psi
(t_{n})\rangle |^{2}$, which means that the ground state we wish to find is
obtained with the probability proportional to $\exp(-\beta(t_n)H_{0})$, since 
$|\Psi (t_{n})\rangle \propto |\Psi_{\mathrm{eq}} (t_{n})\rangle$. If we
carry out the above procedure up to $\beta (t_n)\gg 1$, we can efficiently
obtain the ground state of $H_{0}$. This is called the quantum Jarzynski
annealing (QJA) in the present paper.

It may seem to be unnecessary to apply the time-evolution operator $%
U(\sigma_{k+1}|\sigma_{k};t_{k})$, which expresses the change between states by
quantum fluctuations, at the middle step between the operations of the exponentiated work
 operators $W(\sigma_{k+1};t_{k})$. The time-evolution operator
does not mean an artificial control but describes the change by quantum nature
during quantum computation. Let us remember the nontrivial point of JE. Even
if we allow transitions between the exponentiated work, JE holds as in Eq. (\ref{CJE}).

We emphasize the following three points. First, the scheme of QJA does not
rely on the quantum adiabatic control. The computational time does not
depend on the energy gap. Therefore QJA does not suffer from the energy-gap
closure differently from the ordinary QA. It is thus important to estimate the
required computational cost from the number of the unitary gates for the implementation of QJA as
will be discussed below. Second, from a point of JE, the result is
independent of the schedule to tune the parameter, $\tau $, in the above
manipulations. Third, we do not need the repetition of the pre-determined process to deal with all fluctuations in the nonequilibrium-process average as in the ordinary JE, since the classical ensemble is mapped to the quantum wave function. 
We operate the above procedure to a single quantum system in principal. 
Notice that, since we need a kind of practical techniques to realize QJA, several-time repetitions of experiments should be demanded since the result by quantum measurement should be probabilistic. 
However we should emphasize that this point is not related with the theoretical property of JE attributed to rare events, necessity of all the realizations during the nonequilibrium process, but it comes from quantum nature.

\begin{figure}[tbp]
\begin{center}
\includegraphics[width=60mm]{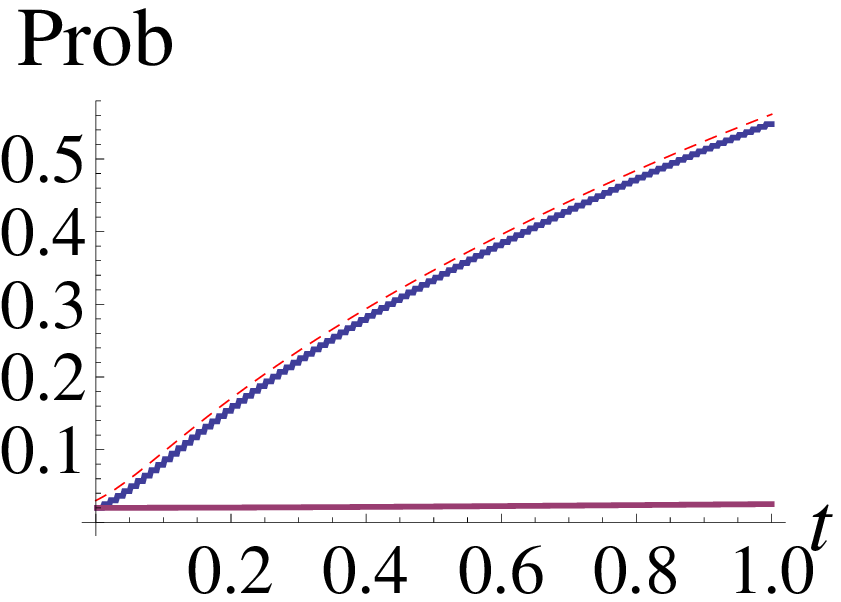} %
\includegraphics[width=60mm]{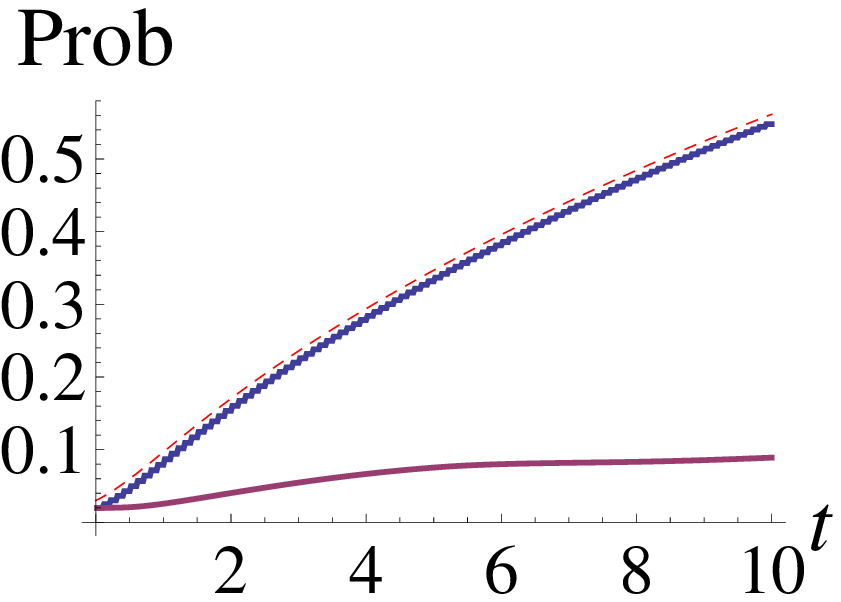} %
\includegraphics[width=60mm]{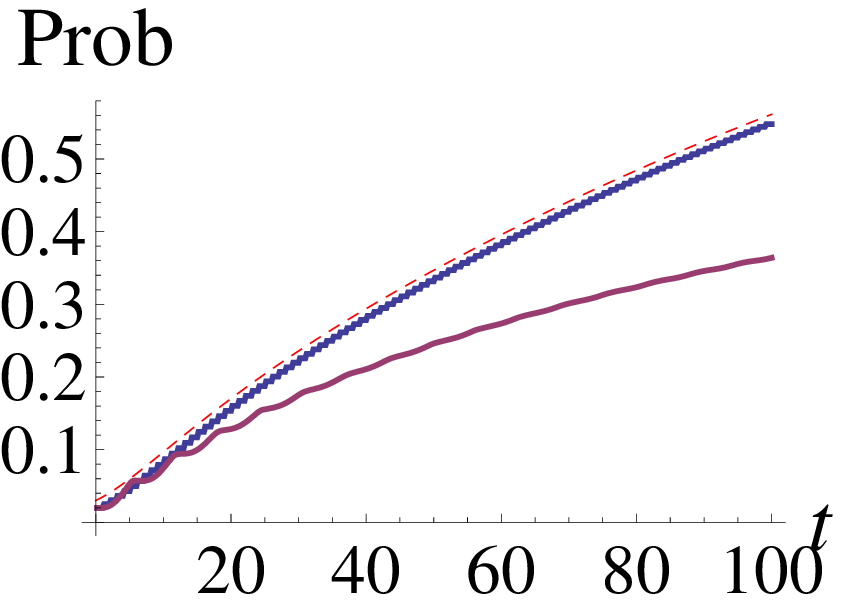}
\end{center}
\caption{{\protect\small (color online) Quantum Jarzynski annealing and
quantum annealing for the one-dimensional random-potential problem. The
probabilities for obtaining the ground state by both of the methods are
plotted for $\protect\tau =1,10$, and $100$ from top to bottom. The dashed
curves denote the instantaneous Gibbs-Boltzmann factor for reference. The
upper solid curves (blue curves) representing the results by QJA are fixed
to these reference curves, while the lower ones (red ones) express the
time-dpendent results by the ordinary QA.}}
\label{QJARP}
\end{figure}

According to the property of JE, we can expect that QJA finds the ground state following the Gibbs-Boltzmann factor independently of annealing schedules.
In contrast, without the multiplication of the exponentiated work, slow quantum control is necessary to efficiently find the ground state according to the ordinary QA. 
Let us take a simple instance to search the minimum from a one-dimensional random potential, which is formulated as the Hamiltonian $H_{0}=-\sum_{i=1}^{N}V_{i}|i\rangle \langle i|$. Here $V_{i}$ denotes the potential
energy at site $i$ and chosen randomly. By the linear schedule for tuning the
parameter $\beta $ from $0$ to $100$, we apply QA without exponentiated work
operations and QJA to the above system. Figure \ref{QJARP} shows the
comparison between the probability for finding the ground state with $N=50$
sites by QA and QJA with different schedules $\tau =1,10$ and $100$. 
%-------------------------------------
%-------------------------------------
The plots for QJA (upper curves) are fixed along the reference curves
(dashed curves) representing the instantaneous Gibbs-Boltzmann factor. In
other words, QJA does not depend on $\tau $, which characterizes the
schedule of quantum computation. In contrast, QA (lower curves) needs
sufficiently slow decrease of quantum fluctuations to efficiently find the
ground state.

To perform QJA, we need to implement the exponentiated work operation $%
W(\sigma _{k};t_{k})=\exp (-(\beta(t_{k+1})-\beta (t_{k}))H_{0}(\sigma
_{k})/2)$, which looks like a non-unitary operator. To implement this
operation, we consider a quantum state with an ancilla qubit (another
two-level quantum system) as $|\Psi ,\phi_1 \rangle =|\Psi \rangle \otimes
|\phi_1 \rangle $, where $\phi_1$ is assumed to take $0$ and $1$ \cite{WQ}.
Initially we set $|\Psi ,\phi_1=0 \rangle$, which is called the computational state below.
It is convenient to assume the case that $H_0(\sigma) > 0$ for any states.
Let us define the following ``unitary" operator for the enlarged quantum
system as 
\begin{eqnarray}
W_{\mathrm{unit.}} &=& \sum_{\sigma }|\sigma \rangle \langle \sigma |\otimes
\left( 
\begin{array}{cc}
\sqrt{y(\sigma )} & \sqrt{1-y(\sigma )} \\ 
-\sqrt{1-y(\sigma )} & \sqrt{y(\sigma )}%
\end{array}%
\right)  \notag \\
&\equiv& I_{\sigma} \otimes Y_1,
\end{eqnarray}
where $y(\sigma )=\exp (-\delta \beta H_{0}(\sigma))$. 
We can obtain the weighted quantum system by applying this operator to the computational state as $\sqrt{y(\sigma )}|\Psi ,\phi_1 =0\rangle $. 
In that sense, we can regard $W_{\mathrm{unit.}}$ as the exponentiated work operation $W(\sigma_{k};t_{k})$ for the quantum state $|\Psi ,\phi_1 =0\rangle $ as above shown for the case in which we increase $\beta$ monotonically. 
We can explicitly evaluate each probability amplitude of $W_{\mathrm{unit.}}|\Psi ,\phi_1=0\rangle $ as 
\begin{eqnarray}
\langle \Psi ,0|W_{\mathrm{unit.}}|\Psi ,0\rangle &=&\sqrt{y(\sigma )} \\
\langle \Psi ,1|W_{\mathrm{unit.}}|\Psi ,0\rangle &=&\sqrt{1-y(\sigma )}
\end{eqnarray}
When we consider measurements of the quantum state, $\sqrt{1-y(\sigma )}|\Psi,\phi_1 =1\rangle $ is regarded as an undesired error state in our computation.
We have to bound the error probability $p_{\mathrm{error}} = 1-y(\sigma )$. 
To decrease the error probability and to avoid negative numbers in the square root, we here demand $p_{\mathrm{error}} \sim \delta \beta \mathrm{max}_{\sigma}H(\sigma)\ll 1$.

In QJA, to gain the relevant weight for the ground state of $H_{0}$, we have
to increase a parameter corresponding to the inverse temperature up to $\beta(t_{n})\epsilon \sim 1$, where $\epsilon $ is the minimum energy gap of the ``classical" Hamiltonian $H_{0}$ (usually given by the energy unit).
Therefore the step of QJA, which corresponds to the step number of the work operation $W_{\mathrm{unit.}}$, is necessary up to $n\equiv \beta(t_{n})/\delta \beta \sim 1/\epsilon \delta \beta $. As a result, the
computational time (the step number of the exponentiated work operation) should become longer as $n\sim 1/\epsilon \delta \beta =\mathrm{max}_{\sigma}H_{0}(\sigma)/\epsilon p_{\mathrm{error}}$ to make the error probability $p_{\mathrm{error}}$ lower in our strategy. 
However the computational time for QJA does not depend on the detailed structure of the cost function.

Since $\delta \beta $ is bound, we have to prepare an enlarged quantum state with $n$ ancilla qubits as $|\Psi ,\phi _{1},\cdots ,\phi _{n}\rangle $ to obtain the quantum state with the relevant weight after $n$-step exponentiated work operation as detailed below. 
The computational state of QJA in this case is $|\Psi ,0,\cdots ,0\rangle $. 
The other states as $|\Psi,1,0\cdots ,0\rangle $, $|\Psi ,0,1,\cdots ,0\rangle $, etc. such that
several ancilla qubits are flipped as $\phi _{i}=1$ are regarded as the error states similarly to the above simple case. 
To gain the weight up to $\exp(-\beta (t_{n})H_{0})$, we consider the $n$-step exponentiated work operations as $I_{\sigma }\otimes Y_{1}\otimes I_{2}\otimes \cdots \otimes I_{n}$, $I_{\sigma }\otimes I_{1}\otimes Y_{2}\otimes I_{3}\otimes \cdots \otimes
I_{n}$, $\cdots $ and $I_{\sigma }\otimes I_{1}\otimes \cdots \otimes
I_{n-1}\otimes Y_{n}$, where $I_{j}$ denotes the identity matrix. 
We then obtain the desired state $|\Psi ,0,0\cdots,0\rangle $ after measurements with the weight as $\exp (-\beta (t_{n})H_{0})=(1-p_{\mathrm{error}})^{n}$. 
The weights for the other states, the error states, are given as $p_{\mathrm{error}}(1-p_{\mathrm{error}})^{n-1}$ for $|\Psi ,1,0,\cdots ,0\rangle $ and $p_{\mathrm{error}}^{2}(1-p_{\mathrm{error}})^{n-2}$ for $|\Psi ,1,1,0,\cdots ,0\rangle $ and so on. 
Therefore we can obtain the desired state by repetition of the experiments when we consider the realistic implementation of QJA in quantum computation. 
The demanded number of the repetition of the same experiments is evaluated as $1/(1-p_{\mathrm{error}})^{n}\sim 1+\mathrm{max}_{\sigma }H_{0}(\sigma )/\epsilon $, which does not depend on the choice of $p_{\rm error}$.

We here summarize the results of the above estimations. 
We tune the value of $p_{\mathrm{error}} \ll 1$ (for instance, $p_{\mathrm{error}} = 0.01$) in order to efficiently yield the desired quantum state as $\exp(-\beta(t_n) H_0)|\Psi,0,\cdots ,0\rangle $.
Simultaneously the computational time for QJA is determined as $n\sim \mathrm{max}_{\sigma }H_{0}(\sigma )/\epsilon p_{\mathrm{error}}$  (in the case $p_{\rm error}=0.01$, $n\sim 100 \mathrm{max}_{\sigma }H_{0}(\sigma )/\epsilon$).
Also the number of the repetition of the same experiments can be estimated as $\sim 1 + \mathrm{max}_{\sigma }H_{0}(\sigma)/\epsilon $. 
Even if the maximum value of the cost function becomes larger by increase of the system size as $\mathrm{max}_{\sigma}H(\sigma)/\epsilon = N^{r}$ where $r$ is an arbitrary positive value, both of the computational time and the number of the ancilla qubits do not diverge exponentially, since $n \sim N^r/ p_{\mathrm{error}}$.
The repetition of the experiments can also be reduced to a moderate value as $\sim 1 + N^r$.

We consider an application of JE to quantum computation as QA to solve the
optimization problems by using the classical-quantum mapping. The
classical-quantum mapping enables us to imitate pseudo-thermal processes in
quantum computation. As we expected, this protocol keeps the quantum system to express
the equilibrium state for the instantaneous inverse temperature. 
To decrease some errors occurring after the exponentiated work operation and measurements, we can not
increase rapidly the inverse temperature to obtain the ground state and we
need additional qubits. Nevertheless the cost for the realization of QJA in
quantum computation does not diverge exponentially, which is the essentially
different point from the ordinary QA and other quantum algorithms. The key
point of QJA is that we need another resource like a \textquotedblleft
memory" in quantum computation instead of cut \textquotedblleft time".
Fortunately, the amount of necessary memory (ancilla qubits) as well as the
computational time for implementation of QJA does not diverge exponentially
by the increase of the system size $N$. Thus the results by QJA shown here
imply that we may overcome the difficulties in hard optimization problems
and solve them in a reasonable time. The present results are preliminary but
we should clarify the efficiency for several interesting hard problems we wish to solve in the future study \cite{MO}. 
We hope that QJA becomes one of the basic algorithms using the quantum nature.

The fruitful discussions with Y. Sughiyama, H. Nishimori, Y. Shikano, S.
Tanaka, S. Miyashita, and S. Morita are acknowledged. This work was
supported by CREST, JST.

%%%%%%%%%%%%%%%%%

\end{document}